# An Effective Payload Attribution Scheme for Cybercriminal Detection Using Compressed Bitmap Index Tables and Traffic Downsampling

S. Mohammad Hosseini, Amir Hossein Jahangir

*Abstract*— Payload Attribution Systems (PAS) are one of the most important tools of network forensics for detecting an offender after the occurrence of a cybercrime. A PAS stores the network traffic history in order to detect the source and destination pair of a certain data stream in case a malicious activity occurs on the network. The huge volume of information that is daily transferred in the network means that the data stored by a PAS must be as compact and concise as possible. Moreover, the investigation of this large volume of data for a malicious data stream must be handled within a reasonable time. For this purpose, several techniques based on storing a digest of traffic using Bloom filters have been proposed in the literature. The false positive rate of existing techniques for detecting cybercriminals is unacceptably high, i.e., many source and destination pairs are falsely determined as malicious, making it difficult to detect the true criminal. In order to ameliorate this problem, we have proposed a solution based on compressed bitmap index tables and traffic downsampling. Our analytical evaluation and experimental results show that the proposed method significantly reduces the false positive rate.

## I. INTRODUCTION

Network forensics is a subset of digital forensics, which deals with crimes in computer networks. It contains a set of traffic recording and analysis techniques used to acquire legal evidence for the purpose of proving or rejecting an accusation. While network security and its tools protect a network from attacks and intrusions, network forensics gathers evidence of cybercrimes. The appearance of new vulnerabilities, possible insider treachery, and the rapid development of network technology means that security tools cannot provide absolute security against threats. Hence, network forensic tools complement intrusion detection and prevention systems to support post-mortem investigations [1], [2].

One important technique of network forensics is "*payload attribution*." In this context, "*attribution*" is the problem of determining the source and destination of traffic instances [3]. An instance of traffic, which is simply called an "*excerpt*," can be a single, multiple, a fraction, or any other combination of packet payloads. A payload attribution system (PAS) captures and records network traffic for an extended duration, i.e., for as long as possible. At the investigation time, a specific excerpt is given to the PAS, and the system determines the source and destination of all packets that carried the excerpt. This system can be used as the core of a network forensic system to investigate cybercrimes committed through computer networks. For example, it could be used in an organization to trace the spread of worms in order to detect infected systems and the source of attack by looking for the signature of the worm in the traffic history. As another example, the system can be used to discover the insider who has disclosed sensitive information of the organization.

The simplest way to develop a PAS is to record and archive raw network traffic. Many traffic recording tools have been proposed so far. See for instance [4], [5]. They can capture and archive full network traffic for high-bandwidth links. These tools, which are somehow similar to surveillance cameras, record all packets transferred via the network link for a predefined duration and retrieve and provide them for security experts and investigators when needed. However, the application of this technique is very limited because of the large volume of required storage. Even if this large storage can be supplied, the determination of source and destination of the excerpt within this huge volume of data would be a challenging problem. Another important problem of traffic recording is the violation of privacy. By using the raw network traffic that is recorded in this approach, it is possible to access personal information of users. The privacy problem makes a traffic recorder inappropriate for most situations. The paradoxical requirements of privacy protection and network forensics are so challenging that a new set of architectures and protocols have been proposed for the Internet [6], [7]. However, they are far from being accepted in the real world.

Kulesh et al. [3], [8] have presented a new PAS which tries to resolve the problems associated with traffic recording systems. The main approach is to store a digest for each packet instead of the whole packet. Traffic packets are digested by a data structure called "*Bloom filter*." The Bloom filter is stored in a permanent storage system in order to be used at the investigation time. Consequently, the stored traffic data is considerably reduced; for example, in a typical

Submitted: 13 Jan. 2017. The authors are with the Department of Computer Engineering, Sharif University of Technology, Tehran 145888-9694, Iran (e-mail: smhoseini@ce.sharif.ir; jahangir@sharif.ir, Corresponding author).

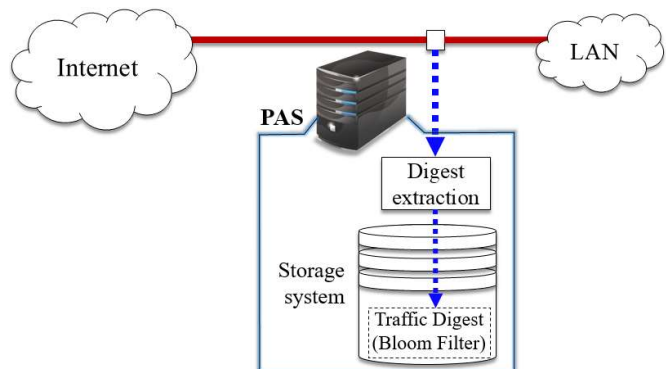

Fig. 1. The general layout of a PAS (Payload Attribution System)

situation, a 4 GB piece of traffic can be digested by a 40 MB Bloom filter (i.e., a data reduction ratio of 100:1). The digesting process of the PAS has a low-time complexity due to the use of the Bloom filter, and consequently, it can digest high-bandwidth traffic in a real-time manner. The impossibility of recovering the traffic from the digest is another feature of this solution which guarantees privacy. Fig. 1 represents the general layout of the PAS and its common deployment location.

The investigation of a malicious excerpt in the PAS is composed of two steps as shown in Fig. 2. In the first step, which is called "*appearance check*" step, the system determines whether or not the excerpt has appeared in a specific time interval of the traffic. In the second step, called "*flow determination*" step, the system determines all the flows (source and destination pairs) that carried the excerpt if the answer to the first step is positive. However, both steps suffer from "*false positive*" issue. A false positive event occurs during the appearance check step if an excerpt does not belong to the traffic however, it is falsely determined as so. A false positive in the flow determination step is a flow that has falsely been reported as the carrier of the excerpt. The false positive rate of the system can be traded off against its data reduction ratio, i.e., a higher data reduction ratio results in a higher false positive rate.

Kulesh reported a relatively low false positive rate for a data reduction ration of 39:1 [3]. Additional studies have sought to improve the false positive rate and the data reduction ratio [9]–[14]. In these works, only the appearance check step has been evaluated, and no report exists on the false positives of the flow determination step. This is a common shortcoming of previous studies while the goal of a PAS is to detect the source and destination of cybercrimes. Evaluations in the previous works did not reveal high false positive rates of the proposed methods. Only under certain conditions, the false positive rate of the appearance check step can be generalized to the flow determination step. Further details on this shortcoming are presented in Section II.

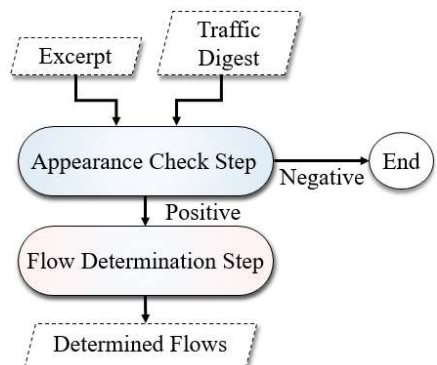

Fig. 2. The investigation procedure of a PAS.

According to recent surveys, the storage and privacy problems are still open challenges in network forensics [15], [16]. Our analysis and experiments, presented in this paper, show that the previous payload attribution systems result in a high false positive rate for the flow determination step; hence, many flows are falsely reported as the excerpt carriers. This problem indicates it is practically infeasible to detect the actual source and destination of the excerpt, because the many false flows that are reported as source and destination pairs make the investigation of the malicious activity confusing. Therefore, these systems have not yet been employed in practice. Our main goal in this paper is to significantly reduce falsely determined flows, and improve the feasibility of payload attribution systems for cybercriminal detection. We believe that if the false positive rate is reduced, the system can also be used in other related fields such as cloud forensics and P2P networks which deal with a huge volume of data [17], [18].

In this paper, first, we propose a downsampling technique which removes parts of traffic less informative for a PAS. This novel technique decreases the insertion rate of Bloom filter and consequently the false positive rate of the PAS. Second, we focus on the flow determination step which has not been deeply studied by the previous works. We present a simple, though effective feature extraction approach, based on the combination of a Bloom filter with a compressed bitmap index table in order to prevent the PAS from exhaustively querying in the flow determination step. As a result of the flow query rate reduction, the false positive rate of the flow determination step will significantly decrease. The idea behind the solution is based on previous research on characteristics of flows in network traffic. We call our PAS "CBID" which is the abbreviation of "Compressed Bitmap Index and traffic Downsampling." We prove the effectiveness of our solution with an information theoretic approach.

The rest of this paper is organized as follows: Section II briefly discusses related works. The proposed PAS is presented in Section III, and Section IV evaluates it. Section V presents some discussions about the proposed method. We conclude the paper in Section VI.

## II. RELATED WORK

All the previous payload attribution systems based on traffic digesting have used Bloom filter as their main building block. In the following, we review the Bloom filter and briefly discuss the evolution of payload attribution systems.

### A. Bloom Filter

A Bloom filter [19] is a space-efficient probabilistic data structure that represents a set in order to support membership queries. It consists of an $m$-bit array and a set of $k$ independent hash functions $h_0, \ldots, h_{k-1}$ ranging from $\{0, \ldots, m-1\}$. All the array entries are initially set to 0. The $n$-member set $S = \{s_0, \ldots, s_{n-1}\}$ is stored in the Bloom filter by setting bits at position $h_i(s_j)$ to 1 for all $0 \leq i \leq k-1$ and $0 \leq j \leq n-1$. A membership query for element $x$ is done by determining whether or not bits at position $h_i(x)$ are set to 1 for all $0 \leq i \leq k-1$. If at least one bit is 0, element $x$ is definitely not a member of the set; otherwise, it is considered to be a member of the set. The space efficiency of a Bloom filter is achieved at the cost of the controllable probability of false positive answers as estimated by Equation 1; however, Bloom filters obviously do not have false negatives. A comprehensive survey of Bloom filters and their network applications can be found in [20], [21].

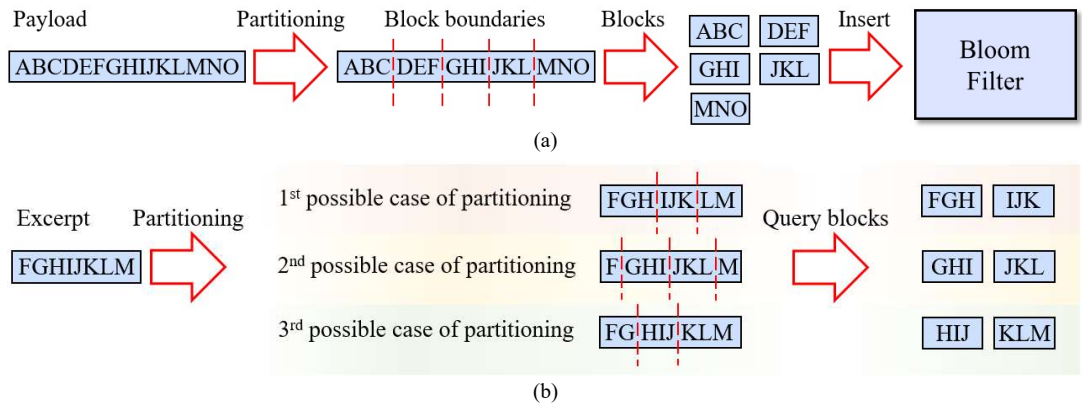

Fig. 3. An example of payload partitioning and excerpt querying (with a partitioning size of 3 bytes).

$$FP = (1 - (1 - \frac{1}{m})^{kn})^k \approx (1 - e^{-\frac{kn}{m}})^k \quad (1)$$

*B. Payload Attribution Systems*

As stated, the simplest way to have a PAS is to store the whole payload of all packets. Kulesh et al. [3] proposed a PAS based on storage of digests of packets in order to decrease the storage volume needed to save the history of traffic. The system, called Hierarchical Bloom Filter (HBF), is capable of answering queries for an excerpt with a low-time complexity and can also preserve privacy. Briefly, HBF works as follows: In the digest generation phase, the payload of each packet is partitioned into equal-sized blocks, and each block is inserted into a Bloom filter. For example, if a payload is *ABCDEFJHIJKLMNO* and the partitioning size is 3 bytes, the blocks which are sampled and inserted into the Bloom filter are as represented in Fig. 3 (a). When the false positive probability of the Bloom filter reaches a predefined value, it is moved to a permanent storage system, and next packets are inserted into another Bloom filter. Consequently, each stored Bloom filter is the traffic digest of a specific time interval.

During an investigation, a query for the appearance check of an excerpt is processed as follows: First, the excerpt is partitioned. Since the excerpt may not start exactly on a block boundary, partitioning is done with all possible alignments. In the above example, if excerpt *FGHIJKLM* is queried, it is partitioned according to the three cases shown in Fig. 3 (b). For each case of partitioning, the stored Bloom filters are queried for all the complete blocks (3-byte blocks in this example). The excerpt will be considered as belonging to the traffic if queries for all blocks of a partitioning get positive answers. In the example, the query will definitively receive a positive answer for the second case for which blocks GHI and JKL are queried. However, the two other cases could also get false positive answers because of the false positive possibility of Bloom filter. HBF uses offsets associated with the blocks in order to verify that the blocks have appeared consecutively in a single payload. Further details can be found in [3].

Up to this point, HBF has only been able to check whether or not a specific excerpt has appeared in the traffic; however, it cannot determine the corresponding source and destination. To address this problem, in addition to the mentioned blocks, HBF stores another type of blocks which are a combination of packet pieces and the flow identifier of the packet. HBF also stores a list of all flows seen during traffic processing. Fig. 4 depicts this process and the two types of blocks, which we call "type-I" and "type-II" blocks. Consequently, there will be a Bloom filter and a flow list for each specific time interval in the storage system. Hence, the investigation of an excerpt proceeds in the following manner: The appearance check step determines the Bloom filters containing the queried excerpt. Then, in the flow determination process, each determined Bloom filter is queried for type-II blocks using all the flows saved in its corresponding flow list. Each flow for which all the blocks of the excerpt receive a positive answer is considered to be a carrier of the excerpt.

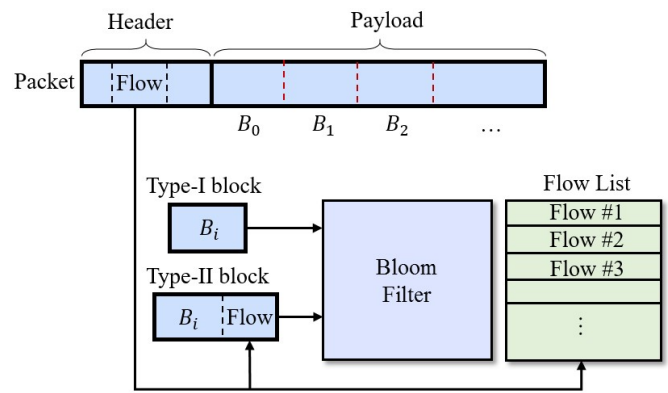

Fig. 4. The process of traffic digest generation. $B_0, B_1, B_2, \ldots$ are pieces of the payload after partitioning. The type-I and type-II blocks are used for appearance check and flow determination, respectively.

Cho et al. [10] proposed a similar approach called rolling Bloom filter (RBF) to resolve the problem of block alignment. The evaluation results of RBF are similar to the best case of HBF as noted in [9] and [12]. Ponec et al. [9], [11] studied the weaknesses of HBF and offered WBS and WMH methods to decrease the false positives. WBS stands for Winnowing Block Shingling. The shingling approach replaces the offsets in the HBF. It uses block overlapping in order to verify the consecutiveness of blocks in a single payload. To remove the need for testing all possible alignments for the first block, WBS does not use fixed-size blocks, but instead sets block boundaries based on computations on payload contents which results in variable-sized blocks. The boundary selection scheme is based on the winnowing fingerprinting algorithm [22]. WMH uses several instances of WBS with different parameters to decrease the false positives. These two methods have been evaluated by querying for the appearance check of excerpts which had not appeared in the experimented traffic. Based on YES/NO answers to these queries, a false positive rate has been reported for each method. Although they reported low false positive rates, our experiments showed unacceptably higher false positive rates. Since they did not evaluate the methods in the flow determination step, they presumably had not inserted the type-II blocks into Bloom filters, which could increase the false positive rate of Bloom filters. The false positive rate of the appearance check step can be generalized to the flow determination step only if the type-II blocks are also inserted into Bloom filters. However, this is not clear in the paper. Our evaluations in Section IV clarify this issue.

Haghighat et al. [12] presented the CMBF method, which supports wildcard queries. This kind of query is useful when only parts of the excerpt are known, such as the signature of a polymorphic worm. In this case, the unknown parts are filled with wildcard bytes. Previous methods must carry out an exhaustive search for these queries which can take a long time and may generate many false positives because of the large number of queries. CMBF has been evaluated similarly, i.e., for the appearance check step and without considering the type-II blocks. Its false positive rate is slightly lower than WBS and WMH for regular queries (without wildcards).

The last proposed method is WDWQ [13] which is based on a combination of WBS and CMBF. It has yielded a lower false positive rate in the same evaluation procedure. However, as our more elaborated experiments show in Section IV, the actual false positive rate of the previous methods is very high. In the next section, we present our new method, CBID, which significantly decreases the false positive rate.

## III. Proposed Method

There are two important reasons for the high false positive rate of a PAS. The first is the high insertion rate of payload blocks which increases the false positive rate of Bloom filters. In order to alleviate this problem, we present a downsampling approach on payload data to decrease the number of blocks inserted into Bloom filters. The second reason is the large number of queries to Bloom filters in the flow determination step. As stated before, after the appearance check step and determination of the Bloom filters containing the queried excerpt, the PAS exhaustively queries for all traffic flows corresponding to the determined Bloom filters. Consequently, the false positive rate of the flow determination step increases with the number of queries to Bloom filters. This important problem has not been addressed by the previous works. We resolve it by a simple feature extraction approach based on Bitmap index tables.

### A. Downsampling

We use winnowing to select block boundaries, and shingling to resolve the consecutiveness problem as discussed in [9]. This scheme samples blocks in a variable-size manner and results in a uniform distribution of blocks as a function of block size. We examined this property by applying the boundary selection scheme on 100 GB of our CE Department network traffic[1], and storing its digest in a Bloom filter with a data reduction ratio of 100:1. The winnowing window and the shingling overlap length of our experiment are 64 and 4 bytes, respectively. Therefore, block sizes must be between 6 and 69 bytes. Fig. 5 shows the distribution of blocks as a function of block size in the experiment.

As can be seen in Fig. 5, the distribution is uniform. The initial peak for the smallest block size is due to low-entropy payloads, such as a long string of zeroes. A similar result has been observed in [9]. The fact that the number of small blocks is approximately equal to the number of larger blocks is the downside of the winnowing technique as a variable-size sampling scheme. Small blocks increase the block insertion rate and consequently, augment the false positive rate of Bloom filters. Fixed-block size schemes, such as HBF and CMBF, have a lower block insertion rate. However, unlike a fixed-block size scheme, the winnowing technique does not suffer from the alignment problem.

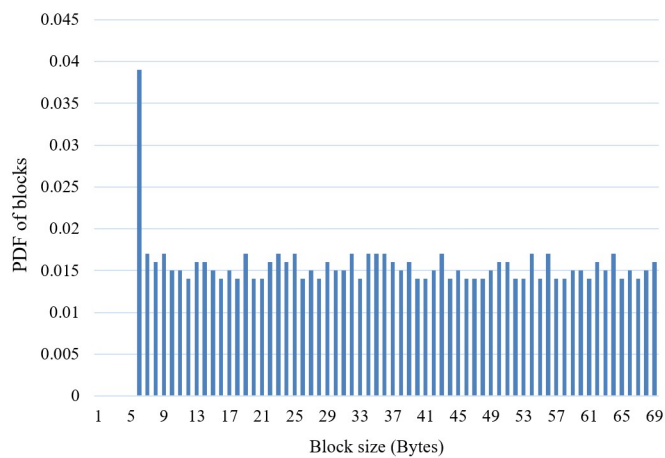

Fig. 5. Distribution of blocks as a function of block size.

[1] We recorded only TCP and UDP packets. The traffic comprises 81% TCP and 19% UDP. The traffic distribution was uniform at recording time. The distribution of traffic flows as a function of flow size is shown in Fig. 8.

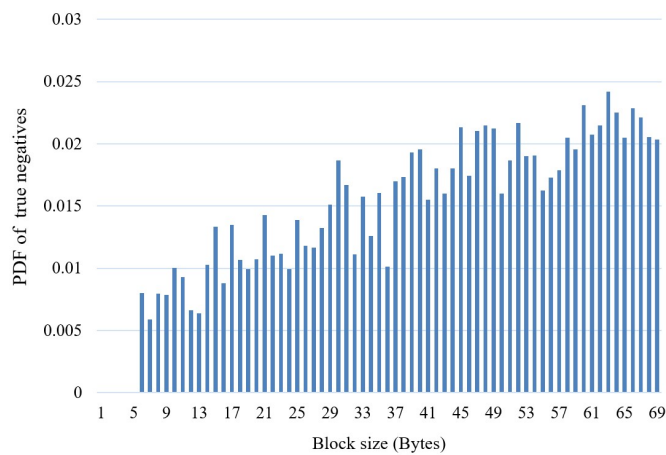

Fig. 6. Distribution of true negative answers as a function of block size.

We exploit the uniform distribution of blocks in our downsampling approach. Obviously, small blocks contain less information than large blocks. Moreover, the sample space of small blocks is evidently smaller than large ones. Hence, in comparison to large blocks, a considerable portion of the sample space of small blocks is inserted into Bloom filters. This decreases the **true negative** probability for small blocks and consequently, makes them less effective in the reduction of false positives. In order to experimentally evaluate it, we queried the Bloom filter for 10000 excerpts, none of which had appeared in the trace. The size of excerpts was 200 bytes. Fig. 6 shows the distribution of true negative answers to queries for blocks of the excerpts. The distribution is shown as a function of block size. Although the distribution of blocks as a function of block size is uniform (Fig. 5), larger blocks are more effective as indicated by Fig. 6. Hence, we propose to downsample payload blocks in the digesting procedure by discarding blocks smaller than a predefined size and inserting only large blocks into Bloom filters. In the investigation time, the answer to a query for such blocks is assumed to be positive. The uniform distribution of blocks guarantees that the participation probability of all sizes in an excerpt are approximately equal and therefore there are enough blocks of large sizes in the excerpt. This scheme not only resolves the alignment problem but also ameliorates the block insertion rate. In other words, we have the advantages of both fixed-size and variable-size schemes. The predefined size below which the blocks are discarded is called "*downsampling threshold*" in this paper. It should also be noted that this scheme does not suffer from false negatives because the answers to queries for blocks smaller than the threshold are assumed to be positive.

From a theoretical standpoint, by using the downsampling technique, the PAS performance is affected by a tradeoff between the false positive probability of the Bloom filter and the number of queried blocks for an excerpt. A larger downsampling threshold reduces the false positive probability of the Bloom filter. On the other hand, a high downsampling threshold decreases the number of blocks of an excerpt that are queried. As evident, a query for an excerpt will not result in a false positive if at least one of its blocks gets a true negative. Therefore, decreasing the number of excerpt blocks for which the PAS queries will increase the false positive probability of the excerpt, i.e., a very high downsampling threshold increases the false positive rate. Hence, the false positive rate of the PAS has a minimum as a function of the threshold. We find the approximate optimum threshold by running several experiments in the evaluation section.

*B. Reducing false flows*

As stated before, since a PAS exhaustively queries for flows in the flow determination step, its false positive rate is high. Reducing the number of flows to be queried can be achieved by extracting features from the result of the appearance check query and attributing them to flows before querying for all of them. This means that the data structure of the Bloom filter should be modified so that some features be returned in case of a positive answer. We use a multi-section Bloom filter for this purpose. In the following, we introduce the multi-section Bloom filter, and then we describe our approach to reducing flow queries in the flow determination step.

*C. Multi-section Bloom Filter*

A multi-section Bloom filter (MSBF) with $j$ sections and a bit array of size $m$ bits, is composed of $j$ smaller Bloom filters and $k+1$ hash functions. For an insertion operation, one hash function is used to select a sub-Bloom filter, and the insertion procedure is done in the selected sub-Bloom filter with $k$ other hash functions. It can be easily proven that the false positive probability of an MSBF with a size of $m$ bits and with $k+1$ hash functions is equal to the false positive probability of a conventional Bloom filter of the same size with $k$ hash functions [23].

The feature that can be returned by an MSBF in case of a positive answer is the section number in which the queried element has been found. Although this feature could be extracted directly by one of the hash functions in a conventional Bloom filter, the MSBF is introduced here due to its additional useful properties. Multi-section Bloom filters have been used by [23] and [24] to implement large high-performance Bloom filters which do not fit into the main memory. Since payload attribution systems need large Bloom filters with a high insertion rate, the MSBF will be an appropriate replacement for the conventional Bloom filter. Therefore, in the proposed solution, we replace conventional Bloom filters with multi-section Bloom filters. However, a compact data structure is needed for storing the features. It should be noted again that the digesting process of a PAS must have a low-time complexity, and the data structure should also follow this rule like Bloom filters. We use a bitmap index table for this purpose.

*D. Bitmap Index Table*

A bitmap index table, which is usually used in database applications, is a data structure that improves the speed of data retrieval operations. In a bitmap index table, a row is considered for each entry, and each row has a bit array (commonly called bitmap) to store features. A bitmap index table is usually queried for some features, and it answers queries by performing bitwise logical operations on the bitmaps and returns the entries that are consistent with those features. Although an index table can increase the speed of a PAS in answering queries, the speed is not our challenging problem. Our aim is to reduce the false positive rate by reducing the flow queries in the flow determination step.

We use a bitmap index table alongside the flow list of each interval. We consider a row having $j$ bits with default values of '0' in the index table for each flow appearing in the time interval. The value of $j$ (number of columns of the index table) is equal to the number of sub-Bloom filters of the MSBF. The digesting process of each traffic flow in the proposed method is as follows:

1. Add the flow into the flow list, and a row in the index table for the flow.
2. Partition payloads of the flow, and extract type-I and type-II blocks.
3. Insert all type-I blocks into the multi-section Bloom filter.
4. In the index table, set the corresponding bit of the sub-Bloom filter into which each type-I block is inserted.
5. Insert all type-II blocks into the multi-section Bloom filter.

For example, if the payload of Flow 1 is composed of 4 blocks and the type-I blocks are inserted into sub-Bloom filters 2, 3, 6 and 10, the bits 2, 3, 6 and 10 of the row corresponding to Flow 1 are respectively set to '1' in the index table, as depicted in Fig. 7; consequently, at the investigation time, we will know there are no type-I blocks related to Flow 1 in the other sub-Bloom filters. In Fig. 7, we know, for example, sub-Bloom filter 6 has type-I blocks from Flow 1, Flow 2 and Flow 4. Therefore, if at least one of the type-I blocks of an excerpt is found in the sub-Bloom filter 6, the system should not query for Flow 3 and Flow 5. This feature can be used to shrink the flow lists before querying flows in the flow determination step and hence, reduce the flow queries. Thus, the investigation process for an excerpt is as follows:

1. Query for the appearance check using type-I blocks, and determine the multi-section Bloom filters (time intervals) into which the excerpt has appeared.
2. For each determined MSBF, determine the sub-Bloom filters into which all the type-I blocks have been inserted.
3. Using the bitmap index table, reject all the flows that have at least a zero bit in the columns corresponding to the determined sub-Bloom filters.
4. Query for flow determination using the remained flows.

For example, in Fig. 7, if the appearance of an excerpt composed of 3 blocks is confirmed in the appearance check step, and the type-I blocks are found in sections 3, 6 and 10 of the multi-section Bloom filter, only Flow 1 and Flow 4, both having a value of '1' in columns 3, 6 and 10, are queried in the flow determination step. In other words, the system generates type-II blocks and queries for them using only Flow 1 and Flow 4. Therefore, the flow queries are reduced.

| Each column corresponds to one section of MSBF | | | | | | | | | | |
|---|---|---|---|---|---|---|---|---|---|---|
| | S1 | S2 | S3 | S4 | S5 | S6 | S7 | S8 | S9 | S10 |
| Flow 1 | 0 | 1 | 1 | 0 | 0 | 1 | 0 | 0 | 0 | 1 |
| Flow 2 | 1 | 0 | 1 | 1 | 0 | 1 | 1 | 1 | 0 | 0 |
| Flow 3 | 0 | 0 | 0 | 1 | 1 | 0 | 0 | 0 | 0 | 0 |
| Flow 4 | 1 | 1 | 1 | 1 | 0 | 1 | 1 | 1 | 1 | 1 |
| Flow 5 | 0 | 0 | 1 | 0 | 1 | 0 | 1 | 0 | 1 | 1 |
| … | | | | | | | | | | |

Fig. 7. An example of the bitmap index table. In this example, blocks of flow 1 are inserted into sections 2, 3, 6 and 10 of the multi-section Bloom filter.

This approach can improve the performance of payload attribution systems if two conditions are met: the first is that the size of the table should be much smaller than the size of the Bloom filter. Obviously, we could increase the size of Bloom filters instead of adding a large data structure. The second condition is that the percentage of bits having a value of '1' in the table should be low. This is important for two reasons: first, it increases the probability of removing false flows; second, it reduces the entropy of the index table.

If the entropy of the index table is low, the table can be highly compressed using methods based on statistical redundancy elimination. This is important for us because we store the index table in a compressed form. It should be noted that a natural strategy to further minimize the storage of the archive contents is to compress the digest before storing it in the permanent storage system. This strategy has been used by some previous works ([3], [9]). At the investigation time, the Bloom filters are decompressed and queried one by one. However, Bloom filters are not capable of being considerably compressed due to their randomized nature (i.e., high entropy). Ponec et al. [9] achieved about 20 percent storage saving by compressing the Bloom filters using GZip. We show that the index table, unlike Bloom filters, has a low entropy and can be highly compressed.

If the percentage of bits having a value of '1' in the index table is low, the index table will contain long strings of zero bits and consequently, low entropy. This factor depends on the number of sub-Bloom filters and the average volume of traffic per flow. The greater amount of data a flow transfers, the more blocks of payloads it will have, and the higher will be the number of bits equal to '1' for the flow in the index table. On the other hand, flows that carry a low volume of traffic will have a fewer number of bits having a value of '1'. Accordingly, shorter flows improve the performance of the index table for eliminating false positives and enhance its ability to be highly compressed.

Based on the characteristics of traffic flows, a small percentage of flows carry the majority of Internet traffic [25]–[31]. The measurements done in previous works show that most of the Internet flows carry individually a low amount of traffic. As a result of this behavior, known as 'mice and elephants' flows, the proposed method is expected to achieve good performance in terms of compression ratio and elimination of false positives.

Before evaluating CBID, we examined the mentioned feature of the traffic in our experiments. Fig. 8 shows the cumulative distribution function (CDF) of flows as a function of flow data size. According to our results, more than 60 percent of traffic flows have carried lower than 2000 bytes. Assuming that each flow within this 60 percent carries on average 1000 bytes, and considering the fact that the average size of blocks is 32 bytes, each flow will have 1000/32=31 blocks, and consequently, 31 bits of the corresponding row in the index table will be set to '1'. Assuming an MSBF with 2048 sections, only (31/2048)×100=1.5 percent of the index table in its 60 percent portion will be set, and therefore the entropy of the table will be very low. The entropy will be even lower by using the downsampling technique. In the next section, we present a detailed analysis and evaluation of CBID.

## IV. Experimental Evaluation

In this section, we evaluate our method and compare it with the previous works. In addition to our method, we implemented the state of the art techniques WBS, WMH [9], CMBF [12] and WDWQ [13] using their most optimized reported parameters[2]. We digested the traffic discussed in Section III using the methods with a data reduction ratio of 100:1, making the digest size 1 GB. In order to evaluate the false positive rate, we extracted 1000 excerpts, each 200 bytes long, while all of them had appeared only once in the traffic. We queried each PAS for determining the carrier flow of each excerpt. The ideal outcome is to report only one flow for each excerpt, i.e., its actual carrier flow. However, there are also false flows due to the Bloom filter false positive probability. We define the false positive rate of a PAS as 'the number of falsely determined flows divided by the total number of distinct traffic flows.' Table I represents the false positive rates. On average, the false positive rates are greater than 10% which means more than 10% of the traffic flows are falsely determined as the carriers of each excerpt. Since the excerpts have appeared only once in the traffic, and each one corresponds to a single flow, the observed false positive rate is very high. Hence, the large number of false flows makes next steps of the investigation and criminal detection impossible.

It is important to note that if we do not insert the type-II blocks into Bloom filters and evaluate the false positive rate of each examined PAS in the appearance check step, we observe a false positive rate very similar to the reported results of the previous works. That is why we believe they have not taken into account the type-II blocks in their evaluation. In the following, we evaluate the downsampling approach and the bitmap index table scheme, and compare them with the previous works. It should be noted that the set of queried excerpts in all experiments are the same.

### A. Evaluation of downsampling

We digested the traffic by our downsampling method using different downsampling thresholds. The winnowing window and the overlap length were 64 and 4 bytes, respectively. Then we queried for the excerpts. The false positive rates are shown in Table I. As explained in the previous section, there is an optimum value for the downsampling threshold due to the tradeoff between the Bloom filter false positive rate and the number of excerpt blocks for which the PAS queries. The proposed approach reduces the false positive rate for downsampling thresholds lower than 40 bytes. For the threshold of 50 bytes, the false positive rate significantly increases since a low portion of excerpt blocks is larger than the downsampling threshold. The 60-byte downsampling threshold results in an excessively high false positive rate. The reason is that some excerpts do not have any blocks larger than 60 bytes, and consequently, they always get a false positive response for each queried flow in the flow determination step. The 40-byte threshold is a proper choice for the optimum threshold which also has a sufficient safety margin from the destructive threshold.

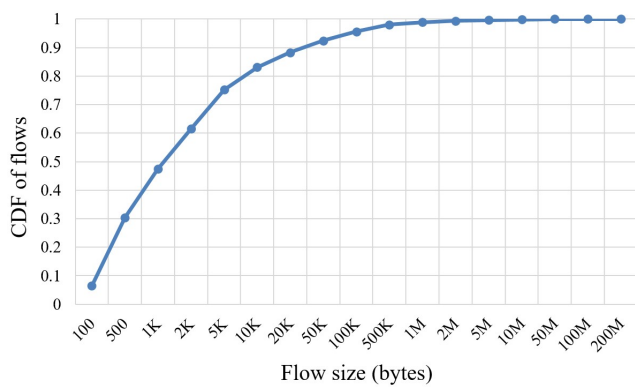

Fig. 8. CDF of traffic flows as a function of flow payload size.

[2] WBS: winnowing window size = 64 bytes, overlap length = 4 bytes. WMH: two instances of WBS. CMBF: hash window size = 8 bytes, aggregation factor = 3, fingerprint modulo = 8. WDWQ: hash window size = 10, fingerprint modulo = 8.

TABLE I: False positive rates for previous methods and the downsampling approach

| Method | | False positive rate |
|---|---|---|
| WBS | | 0.144 |
| WMH | | 0.140 |
| CMBF | | 0.141 |
| WDWQ | | 0.138 |
| Downsampling method | Threshold = 10 | 0.130 |
| | Threshold = 20 | 0.080 |
| | Threshold = 30 | 0.076 |
| | Threshold = 40 | 0.074 |
| | Threshold = 45 | 0.075 |
| | Threshold = 50 | 0.391 |
| | Threshold = 60 | 0.966 |

TABLE II: Statistical analysis of bitmap index table

| MSBF sections | 1024 | | | | 2048 | | | | 4096 | | | | 8192 | | | |
|---|---|---|---|---|---|---|---|---|---|---|---|---|---|---|---|---|
| Size of index table (MB) | 108 | | | | 216 | | | | 432 | | | | 864 | | | |
| Bits having a value of '1' (%) | 13.1 | | | | 9.5 | | | | 6.0 | | | | 4.1 | | | |
| Symbol size (Bytes) | 8 | 32 | 128 | 256 | 8 | 32 | 64 | 128 | 8 | 16 | 32 | 64 | 8 | 16 | 32 | 64 |
| Entropy (bits) | 15.3 | 24.0 | 27.9 | 27.5 | 13.1 | 23.0 | 26.8 | 26.9 | 10.8 | 19.0 | 22.7 | 24.0 | 9.0 | 11.9 | 12.7 | 13.3 |
| Best theoretical compression ratio | 3.0 | 3.5 | 4.0 | 3.9 | 4.9 | 6.1 | 6.3 | 6.2 | 6.0 | 6.5 | 7.0 | 6.8 | 7.2 | 7.7 | 8.1 | 7.8 |
| Minimum theoretical size of compressed index table (MB) | 30.0 | 27.7 | 26.3 | 27.0 | 44.1 | 35.4 | 34.3 | 34.8 | 72 | 66.5 | 61.7 | 63.5 | 120 | 112 | 107 | 111 |
| Size of compressed index table using LZMA2 (MB) | 44 | | | | 65 | | | | 98 | | | | 154 | | | |
| Overall data reduction ratio ($DR_O$) | 96 : 1 | | | | 94 : 1 | | | | 91 : 1 | | | | 87 : 1 | | | |

## B. Evaluation of Bitmap index table

We modified the system described in the previous subsection to use the MSBF architecture instead of the conventional Bloom filter. The system creates the bitmap index table for flows at digesting time and utilizes it in the investigation procedure. Before evaluating the impact of the bitmap index table on the false positive rate, we evaluated the percentage of bits having a value of '1', the entropy and the maximum theoretical amount of compression of the index table as follows: first, we digested the traffic with a data reduction ratio of 100:1, and an index table was created for the flows. The downsampling feature was disabled in this experiment. Then, we examined the percentage of bits having a value of '1', the entropy and the maximum theoretical amount of compression for the index table. In this experiment, the entropies were calculated based on different symbol sizes. Moreover, the size of the codeword list was considered in the computation of the size of the compressed index table. The results of this experiment, which was run for different segmentations of the MSBF, are shown in Table II. In addition to the maximum theoretical compression ratio derived from entropy, the sizes of index tables compressed using the LZMA2 algorithm are also reported in Table II. Moreover, the overall data reduction ratio of the system ($DR_O$), which is determined by Equation 2, is also reported in this table.

$$DR_O = \frac{\text{Raw traffic size}}{\text{Bloom filter size} + \text{Compressed index table size}}$$

$$= \frac{100\ (GB)}{1\ (GB) + \text{Compressed index tables}\ (GB)} \quad (2)$$

As seen, both the percentage of bits having a value of '1' and the entropy are small. For example, in the case of using an MSBF composed of 2048 sub-Bloom filters, more than 90% of the index table bits are equal to zero. This means that the flow queries are reduced at least by 90%. It should be noted that the number of bits which are checked in the index table per flow is equal to the number of excerpt blocks, and a zero bit for a block is enough to reject the flow; hence, the false positive rate will be reduced by more than 90% in this case. The size of the compressed index table is 65 MB, and therefore, the overall data reduction ratio of the system decreases by approximately 6%. Obviously, the entropy and the compressed index table size will be even smaller by enabling the downsampling feature. This is observed in the following experiment.

## C. Evaluation of CBID

To evaluate the false positive rate, we digested the traffic using different segmentations of MSBF and different downsampling thresholds and then queried for the excerpts. The data reduction ratio was 100:1 when considering only the Bloom filter size as the digest size. Fig. 9 shows the achieved false positive rates. As can be seen, in all configurations of our PAS, the false positive rate significantly decreases in comparison with the results of the previous works shown in Table I; however, the size of the compressed index tables must also be considered prior to the final conclusion.

TABLE III: Overall data reduction ratio ($DR_O$) of CBID

| | | Downsampling threshold | | | | |
|---|---|---|---|---|---|---|
| | | 0 | 10 | 20 | 30 | 40 |
| MSBF sections | 1024 | 96:1 | 96:1 | 96:1 | 96:1 | 97:1 |
| | 2048 | 94:1 | 94:1 | 94:1 | 95:1 | 95:1 |
| | 4096 | 91:1 | 91:1 | 92:1 | 92:1 | 93:1 |
| | 8192 | 87:1 | 87:1 | 88:1 | 88:1 | 89:1 |

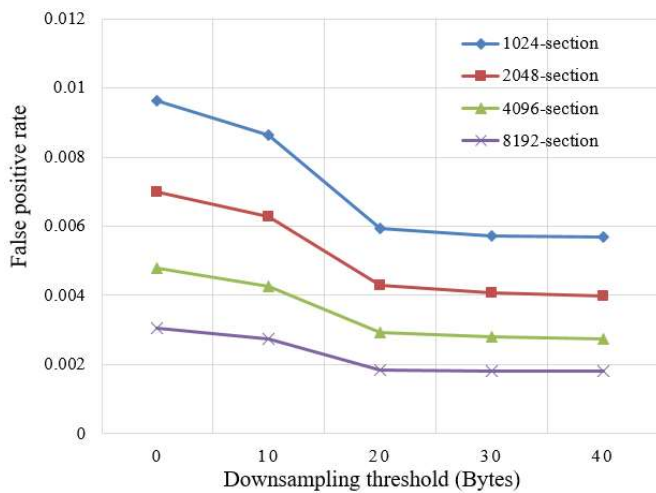

Fig. 9. CBID false positive rate for different configurations

TABLE IV: FALSE POSITIVE RATE OF THE PAYLOAD ATTRIBUTION SYSTEMS AS A FUNCTION OF DATA REDUCTION RATIO

|  |  | Data reduction ratio | | | | | | | | | |
|---|---|---|---|---|---|---|---|---|---|---|---|
|  |  | 97 : 1 | 96 : 1 | 95 : 1 | 94 : 1 | 93 : 1 | 92 : 1 | 91 : 1 | 89 : 1 | 88 : 1 | 87 : 1 |
| Method | WBS | 0.138 | 0.135 | 0.131 | 0.126 | 0.120 | 0.113 | 0.105 | 0.088 | 0.078 | 0.067 |
|  | WMH | 0.134 | 0.130 | 0.125 | 0.120 | 0.115 | 0.108 | 0.100 | 0.081 | 0.069 | 0.059 |
|  | CMBF | 0.135 | 0.131 | 0.125 | 0.120 | 0.116 | 0.109 | 0.102 | 0.082 | 0.069 | 0.060 |
|  | WDWQ | 0.132 | 0.128 | 0.124 | 0.119 | 0.112 | 0.105 | 0.097 | 0.077 | 0.066 | 0.054 |
|  | CBID | 0.006 | 0.006 | 0.004 | 0.004 | 0.003 | 0.003 | 0.004 | 0.002 | 0.002 | 0.003 |

We determined the overall data reduction ratio ($DR_O$) of each configuration as shown in Table III. Then, we repeated the false positive rate evaluation of previous works with the data reduction ratios shown in this table. Table IV represents the false positive rates of all the methods as a function of data reduction ratio. As can be seen in Table IV, CBID outperforms the previous methods for all data reduction ratios. The false positive rate of CBID is on average 22 times less than the previous methods for the case of a data reduction ratio of 97:1. For the data reduction ratio of 89:1, the false positive rate of CBID is lower than 0.002 and may be sufficient for practical investigations. For this data reduction ratio, the false positive rate of CBID is on average 41 times less than the previous methods.

Table V represents the computation time of digesting and querying for each method in our system. We have used one core of a Core i7 CPU with 8 GB of DDR3 RAM and an HDD storage. The reported digesting time is the computation time of digesting the 100 GB traffic using the configurations for the data reduction ratio of 97:1. The querying time is the computation time of querying for one excerpt. No considerable computation time overhead was seen. A detailed discussion on the computation cost is presented in the next section.

V. DISCUSSION

A. Computation cost

In the following, we discuss the impacts of performance-related parameters on the computation overhead. As can be seen in the following, the computation overhead of a PAS is inversely related to its performance, and no optimum value can be found for it.

Table VI represents the computation cost of the methods in detail. The boundary selection cost of CBID is equal to WBS and WDWQ. All of them use winnowing and therefore process a $b$-byte string twice. WMH uses $t$ instances of WBS, and therefore its boundary selection cost is $t$ times more than WBS. CMBF, which uses a fixed-block size scheme, has the least boundary selection cost.

If the winnowing window size of WBS is set as twice the block size of CMBF, the number of blocks which must be processed will be approximately the same. In Table VI, the number of blocks of WBS is denoted as $n$. As a result of the downsampling approach, CBID has the least number of blocks which is equal to $n/d$. The value of $d$ depends on the downsampling threshold. In our experiment, the downsampling threshold of 40 bytes yields $d=2.2$.

The block processing cost for all previous methods is composed of computation cost of $k$ hash functions, and consequently, $k$ memory accesses to insert a block into a Bloom filter. CBID uses $k+1$ hash functions for MSBF and also an additional memory access for the bitmap index table. However, it should be noted that CBID decreases the flow queries in the flow determination step.

Finally, the significant overhead of CBID is the compression cost of bitmap index tables. Bloom filters are stored in a permanent storage system without any additional overhead; nevertheless, bitmap index tables must be compressed before being stored. Since the entropy of a bitmap index table is very low, its compression process does not require high memory footprint and CPU time. It is worth mentioning that in the investigation time, it is not needed to decompress all the stored bitmap index tables. According to the results of the appearance check step, only the bitmap index tables of the time intervals comprising the excerpt will be decompressed.

B. Attacks on PAS

In the following, we discuss some significant ways an

TABLE V: COMPUTATION TIME (MINUTES)

|  | Digesting | Querying |
|---|---|---|
| WBS | 135 | 0.15 |
| WMH | 147 | 0.15 |
| CMBF | 84 | 0.14 |
| WDWQ | 136 | 0.15 |
| CBID | 141 | 0.20 |

TABLE VI: COMPUTATION COST OF THE METHODS

|  | WBS | WMH | CMBF | WDWQ | CBID |
|---|---|---|---|---|---|
| Boundary selection cost | 2b | 2tb | b | 2b | 2b |
| Number of blocks | n | tn | n | 2n | n/d |
| Block processing cost | k | k | k | k | k+2 |
| Storage processing cost | 0 | 0 | 0 | 0 | >0 |

$b$: length of data. $t$: number of WBS instances employed in WMH.
$n$: number of blocks which must be processed by WBS.
$d$: block reduction factor of CBID. $k$: number of hash functions.

adversary can evade the PAS. A more comprehensive discussion about possible attacks on payload attribution systems and solutions for mitigating their effects have been presented in [8], [9].

- Encryption: Just like any other PAS, an excerpt which has been encrypted and transferred through the network, can only be queried by having its encrypted form. A possible solution to overcome this problem is to use payload attribution systems in a distributed manner. In this scheme, encryption algorithms are provided with the digest extraction ability. Each network node, which has access to plain data of its own encrypted traffic, sends digests to the network's central forensics server. Since digests are not reversible to the plain data, privacy is preserved. Moreover, the network traffic workload does not considerably increase due to the low volume of digests. Kulesh et al. [8] have presented such a distributed framework for a PAS, and discussed its deployment challenges.
- Small blocks: The downsampling approach makes CBID vulnerable to payloads composed only of blocks smaller than the downsampling threshold. Moreover, a large number of small blocks can reduce the performance of bitmap index table. However, since the boundaries are selected using winnowing on a hashed payload, it is highly improbable to have payloads comprised of a large number of small blocks.
- Fragmentation: An adversary can divide a malicious data stream among very small packets by using a low TCP segment size. Therefore, the PAS may not be able to generate any block boundaries for the packets. A solution is to make the PAS stateful [8], [9] in which it reassembles payloads of one data stream prior to processing at the cost of additional memory and computation.

## VI. CONCLUSION

This paper presented a brief survey of previous works on payload attribution systems as a tool for storing digests of network traffic to be used in network forensics applications. We proposed a solution based on the combination of Bloom filters and compressed bitmap index tables to reduce the false positive rate of payload attribution systems. Moreover, our scheme has been enriched by a downsampling technique. An analytical approach based on the characteristics of network traffic statistics and information theory showed that the proposed method could significantly decrease the false positive rate without negatively impacting the data reduction ratio. The proposed method was compared with the approaches of previous works using the same data reduction ratio. The results show that the false positive rate decreases at least one order of magnitude. Future studies will seek to add the ability to answer wildcard queries with at least the same achieved false positive rate.

## REFERENCES

[1] G. Palmer, "A Road Map for Digital Forensic Research," in *Proceedings of the 2001 Digital Forensics Research Workshop (DFRWS 2001)*, New York, USA, Aug. 2001, pp. 1–42.
[2] E. S. Pilli, R. C. Joshi, and R. Niyogi, "Network forensic frameworks: Survey and research challenges," *Digital Investigation, Elsevier*, vol. 7, no. 1–2, 2010, pp. 14–27.
[3] K. Shanmugasundaram, H. Brönnimann, and N. D. Memon, "Payload Attribution Via Hierarchical Bloom Filters," in *Proceedings of the 11th ACM Conference on Computer and Communications Security*, Washington, USA, Oct. 2004, pp. 31–41.
[4] L. Deri, A. Cardigliano, and F. Fusco, "10 Gbit line rate packet-to-disk using n2disk," in *Computer Communications Workshops (INFOCOM WKSHPS), 2013 IEEE Conference on*, Toronto, Canada, Apr. 2013, pp. 3399–3404.
[5] V. Moreno, P. M. S. Del Rio, J. Ramos, J. L. G. Dorado, I. Gonzalez, F. J. G. Arribas, and J. Aracil, "Packet Storage at Multi-gigabit Rates Using Off-the-Shelf Systems," in *2014 IEEE International Conference on High Performance Computing and Communications (HPCC 2014)*, Paris, France, Aug. 2014, pp. 486–489.
[6] M. Afanasyev, T. Kohno, J. Ma, N. Murphy, S. Savage, A. C. Snoeren, and G. M. Voelker, "Privacy-preserving network forensics," *Communications of the ACM*, vol. 54, no. 5, May 2011, pp. 78–87.
[7] D. G. Andersen, H. Balakrishnan, N. Feamster, T. Koponen, D. Moon, and S. Shenker, "Accountable Internet Protocol (AIP)," *ACM SIGCOMM Computer Communication Review*, vol. 38, no. 4, Aug. 2008, pp. 339–350.
[8] K. Shanmugasundaram, N. Memon, A. Savant, and H. Bronnimann, "ForNet : A Distributed Forensics Network," in *Second International Workshop on Mathematical Methods, Models, and Architectures for Computer Network Security. Springer*, Sept. 2003, pp. 1–16.
[9] M. Ponec, P. Giura, J. Wein, and H. Brönnimann, "New payload attribution methods for network forensic investigations," *ACM Transactions on Information and System Security*, vol. 13, no. 2, 2010, pp. 1–32.
[10] C. Cho, S. Lee, and C. Tan, "Network forensics on packet fingerprints," in *21st IFIP International Information Security Conference, Springer*, 2006, pp. 401–412.
[11] M. Ponec, P. Giura, H. Brönnimann, and J. Wein, "Highly efficient techniques for network forensics," in *Proceedings of the 14th ACM Conference on Computer and Communications Security (CCS '07)*, Alexandria, USA, Nov. 2007, pp. 150–160.
[12] M. H. Haghighat, M. Tavakoli, and M. Kharrazi, "Payload Attribution via Character Dependent Multi-Bloom Filters," *IEEE Transactions on Information Forensics and Security*, vol. 8, no. 5, 2013, pp. 705–716.
[13] Y. Wei, F. Xu, X. Chen, Y. Pu, J. Shi, and S. Qing, "Winnowing Double Structure for Wildcard Query in Payload Attribution," in *17th International Conference Information Security*, Hong Kong, China, Oct. 2014, pp. 454-464.
[14] Y. Wei, X. Fei, X. Chen, J. Shi, and S. Qing, "Winnowing Multihashing Structure with Wildcard Query," in *The 16th Asia-Pacific Web Conference (APWeb)*, Changsha, China, Sep. 2014, pp. 265-281.
[15] S. Khan, A. Gani, A. W. A. Wahab, M. Shiraz, and I. Ahmad, "Network forensics: Review, taxonomy, and open challenges," *Journal of Network and Computer Applications*, vol. 66, 2016, pp. 214–235.
[16] D. Quick and K. K. R. Choo, "Impacts of increasing volume of digital forensic data: A survey and future research challenges," *Digital Investigation*, vol. 11, no. 4, 2014, pp. 273–294.
[17] V. Roussev, I. Ahmed, A. Barreto, S. McCulley, and V. Shanmughan, "Cloud forensics–Tool development studies & future outlook," *Digital Investigation*, vol. 18, 2016, pp. 1–17.
[18] S. Zawoad, A. K. Dutta, and R. Hasan, "Towards building forensics enabled cloud through secure logging-as-a-service," *IEEE Transactions on Dependable and Secure Computing*, vol. 13, no. 2, March/April 2016, pp. 148–162.
[19] B. H. Bloom, "Space/time trade-offs in hash coding with allowable errors," *Communications of the ACM*, vol. 13, no. 7, pp. 422–426, 1970.
[20] A. Broder and M. Mitzenmacher, "Network Applications of Bloom Filters: A Survey," *Internet Mathematics*, vol. 1, no. 4, 2004, pp. 485–509.
[21] S. Geravand and M. Ahmadi, "Bloom filter applications in network security: A state-of-the-art survey," *Computer Networks*, vol. 57, no. 18,

May 2013, pp. 4047–4064.
[22] S. Schleimer, D. S. Wilkerson, and A. Aiken, "Winnowing: Local Algorithms for Document Fingerprinting," in *Proceedings of the 2003 ACM SIGMOD International Conference on Management of data - SIGMOD '03*, San Diego, USA, Jun. 2003, pp. 76–85.
[23] M. Canim, G. Mihaila, and B. Bhattacharjee, "Buffered Bloom filters on solid state storage," in *Proceedings of the International Workshop on Accelerating Data Management Systems using Modern Processor and Storage Architectures (ADMS)*, Singapore, Sep. 2010, pp. 1–8.
[24] B. Debnath, S. Sengupta, J. Li, D. J. Lilja, and D. H. C. Du, "BloomFlash: Bloom Filter on Flash-Based Storage," in *31st International Conference on Distributed Computing Systems*, Minneapolis, USA, Jun. 2011, pp. 635–644.
[25] W. Fang and L. Peterson, "Inter-AS traffic patterns and their implications," in *Seamless Interconnection for Universal Services. Global Telecommunications Conference. GLOBECOM'99*, Rio de Janeireo, Brazil, Dec. 1999, vol. 3, pp. 1859–1868.
[26] L. Guo and I. Matta, "The war between mice and elephants," in *Ninth International Conference on Network Protocols*, California, USA, Nov. 2001, pp. 180–188.
[27] Y. Zhang, L. Breslau, V. Paxson, and S. Shenker, "On the characteristics and origins of internet flow rates," *ACM SIGCOMM Computer Communication Review*, vol. 32, no. 4, Aug. 2002, pp. 309–322.
[28] K. C. Lan and J. Heidemann, "A Measurement Study of Correlations of Internet Flow Characteristics," *Computer Networks*, vol. 50, no. 1, 2006, pp. 46–62.
[29] T. Benson, A. Akella, and D. a. Maltz, "Network traffic characteristics of data centers in the wild," in *Proceedings of the 10th ACM SIGCOMM Conference on Internet measurement*, Melbourne, Australia, Nov. 2010, pp. 267–280.
[30] P. Hurtig, W. John, and A. Brunstrom, "Recent Trends in TCP Packet-Level Characteristics," in *The Seventh International Conference on Networking and Services, ICNS*, Venice, Italy, May. 2011, pp. 49–56.
[31] X. Zhang and W. Ding, "Comparative Research on Internet Flows Characteristics," in *Third International Conference on Networking and Distributed Computing (ICNDC)*, Hangzhou, China; Oct. 2012, pp. 114–118.

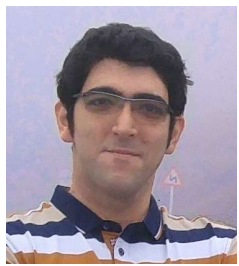

**S. Mohammad Hosseini** received his M.Sc. degree in computer engineering from Sharif University of Technology, Tehran, Iran, in 2014. He has designed and implemented various high-performance network appliances based on FPGAs, specifically, a 40 Gb/s network device tester during his M.Sc. thesis. He is currently a Ph.D. student of computer engineering in Sharif University of Technology. His research interest areas include computer networks, computer architecture and high-performance computing.

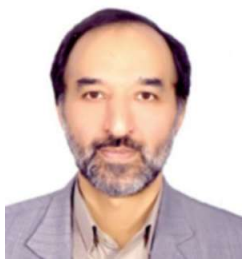

**Amir Hossein Jahangir** obtained his PhD in industrial informatics from Institut National des Sciences Appliquées, Toulouse, France in 1989. Since then he has been with the Department of Computer Engineering, Sharif University of Technology, Tehran, Iran, and has served during his career as the Head of the Department, Head of Computing Center, and is now an associate professor and the director of Network Evaluation and Test Laboratory, an accredited internationally recognized laboratory in the field of network equipment test. His fields of interest comprise network equipment test and evaluation methodology, network security, and high-performance computer architecture.